# Internet and Enterprise Strategy Revisited: From Digital Connectivity to AI-Enabled Enterprises, 1996 - 2026

Carlos J. Costa
*ISEG, Universidade de Lisboa*
Lisbon, Portugal
cjcosta@iseg.ulisboa.pt

***Abstract - This paper revisits the 1996 article "Internet and Enterprise Strategy" and assesses how its propositions were confirmed, transformed, or contradicted as digital connectivity evolved into an AI-enabled enterprise environment by 2026. The reassessment combines historical interpretation with a structured comparison of Internet-enabled capabilities, industry structure, value-chain transformation, platform intermediation, cybersecurity, data governance, artificial intelligence, and digital sovereignty. The analysis shows that the article correctly anticipated the strategic relevance of electronic markets, cybermediaries, transaction security, online communities, and internal rules for Internet use. However, the contemporary Internet exceeds the original framing: it is now a platformized, data-intensive, cloud-dependent, regulated, geopolitically contested, and increasingly AI-mediated environment. The paper argues that this progression from connectivity to AI-enabled enterprise systems does not replace Internet strategy; it extends it. For many firms, enterprise strategy is now enacted through digital infrastructures, algorithmic coordination, ecosystem participation, and governance capabilities.***

## I. INTRODUCTION

When "Internet and Enterprise Strategy" was published in 1996, the Internet was still an emerging business infrastructure [1]. The article framed the Internet as a strategic technology capable of reshaping environmental analysis, industry structure, and the activities of Porter's value chain [2], [3]. At that time, the Web was expanding, electronic commerce was experimental, Portuguese Internet adoption remained limited, and many firms were only beginning to consider whether they needed a Web presence.

Thirty years later, most of the strategic questions raised in 1996 remain relevant, but their scale and implications have changed radically. Digital transformation has moved from peripheral experimentation to a core dimension of enterprise strategy, with artificial intelligence, cloud computing, automation, cybersecurity, platform ecosystems, digital payments, and data governance shaping competitiveness [4], [5], [6].

This paper therefore makes three contributions. First, it provides a retrospective validation of the 1996 propositions in light of contemporary digital business conditions. Second, it explains how enterprise strategy progressed from Internet connectivity and online transactions to platformized, data-intensive, cloud-dependent, and AI-enabled operating environments. Third, it identifies the technical and governance extensions required by this progression, particularly data architectures, AI-enabled decision processes, platform governance, cybersecurity exposure, cloud dependency, and digital sovereignty. The resulting interpretation preserves the original article's focus on Internet-enabled enterprise transformation while showing that competitive advantage now depends less on connectivity alone and more on orchestrating digital resources across organizational, technological, and ecosystem boundaries.

## II. ANALYTICAL APPROACH

The reassessment uses a conceptual, integrative, and historically structured comparison rather than an empirical test. This approach is appropriate when the objective is to synthesize an interdisciplinary body of knowledge, reinterpret an earlier framework in light of subsequent developments, and derive an updated conceptual account rather than estimate causal effects. Consistent with guidance on concept-centric and integrative reviews, the 1996 article provides the baseline framework, while relevant developments documented through 2026 are organized by analytical concepts rather than summarized source by source [33], [34]. Its major propositions are examined across six dimensions: technological infrastructure, market intermediation, value-chain reconfiguration, organizational governance, risk management, and strategic control over data and digital assets. A structured comparison across a common set of dimensions supports systematic interpretation of continuity and change over time [35]. This design also reflects the nature of the original article, which offered a strategic interpretation of how Internet technologies could affect firms rather than a predictive model formulated as measurable hypotheses.

Each dimension is assessed through an explicit three-category coding scheme that makes the comparison transparent and replicable at the conceptual level [33] - [35]. *Confirmation* denotes propositions that became broadly observable in contemporary enterprise practice, such as the normalization of electronic commerce or the integration of the Internet into value-chain activities. *Transformation* denotes propositions that were directionally correct but evolved into more complex forms, as in the shift from disintermediation to platform-based reintermediation. *Extension* denotes strategic layers that were not central to the original article but emerged as connectivity developed into a data-intensive and AI-enabled enterprise environment, including artificial intelligence, data governance, cloud concentration, regulatory compliance, and digital sovereignty. Applying the same categories across all six dimensions provides a consistent basis for distinguishing persistence, modification, and conceptual expansion while avoiding claims of statistical generalizability.

## III. THE 1996 ARGUMENT: MAIN CONTRIBUTIONS

The 1996 article made five main contributions. First, it presented the Internet as a growing infrastructure whose services - e-mail, FTP, Telnet, discussion groups, the Web, and IRC - could alter organizational communication and market interaction. Second, it argued that Internet use should be incorporated into strategic environmental analysis, including technological, sociocultural, economic, legal, demographic, and physical dimensions. Third, it connected information technologies with Porter's structural analysis of industries, showing how the Internet could affect barriers to entry, buyer and supplier power, substitute products, and rivalry [2], [3]. Fourth, it used the value chain to identify how the Internet could support cost reduction and differentiation across infrastructure, human resources, technology development, procurement, logistics, operations, marketing, sales, and after-sales service [7]. Fifth, it highlighted unresolved problems, including copyright protection, payments, security, rules of use, and online communities [8], [9]. Technically, the article was important because it did not reduce the Internet to a marketing channel. It treated connectivity, information exchange, interorganizational coordination, and market access as components of enterprise architecture, even though the language of digital architecture, platforms, cloud services, and AI systems had not yet become common in management research.

## IV. WHAT WAS CONFIRMED BY 2026

**First, the Internet became central to competitive strategy.** The original argument was confirmed. The Internet is now embedded in business models, supply chains, customer interfaces, payment systems, public services, and organizational routines. In 2026, enterprise competitiveness depends not only on Internet access or a website, but on integrating digital capabilities into operating models, data architectures, customer experience, and innovation processes [4], [6].

**Second, the article correctly anticipated the importance of electronic markets and online transactions.** In 1996, electronic commerce was still limited by payment and security constraints. By 2026, e-commerce had become a normal component of retail and B2B exchange, confirming the earlier claim that the Internet would become an economically relevant channel for distribution and transactions [10].

**Third, the analysis of the value chain was strongly confirmed.** The Internet now affects virtually every value-chain activity. Digital marketplaces and supplier portals mediate procurement; logistics depend on real-time tracking and data exchange; operations use cloud computing, automation, analytics, and the Internet of Things; marketing relies on search, social media, personalization, and platform advertising; and after-sales service uses chatbots, remote diagnostics, digital documentation, and online communities. The 1996 paper's proposition that the Internet would reduce communication costs, improve information access, and support differentiation has been widely validated [1], [7].

**Fourth, the emphasis on security and rules was highly prescient.** The 1996 article treated transaction security, copyright, payment reliability, and norms of use as strategic constraints. In 2026, cybersecurity, privacy, identity management, fraud prevention, compliance, and resilience are board-level strategic priorities because firms depend on interconnected cloud services, remote work, digital payments, and data-intensive operations [11], [12].

**Fifth, online communities became strategic knowledge and demand infrastructures.** The 1996 discussion of communities anticipated a major shift from one-way corporate communication to interactive, user-generated, and networked value creation. By 2026, communities influence product discovery, reputation formation, customer support, open-source development, brand advocacy, and innovation feedback loops. Their strategic role is not limited to communication; they function as distributed sensing mechanisms through which firms detect weak signals, monitor complaints, observe emerging use cases, and identify complementary innovation opportunities [8].

**Sixth, transaction security evolved into enterprise-wide cyber resilience.** The original emphasis on secure payments and trust mechanisms has expanded into a broader architecture of identity management, encryption, access control, monitoring, incident response, third-party risk management, and business continuity. In contemporary firms, cybersecurity is not only a technical function but a design constraint for products, data platforms, partner interfaces, and AI-enabled services. The rise of AI increases this complexity because organizations must secure both human users and non-human agents operating across applications and workflows [11], [13], [14].

## V. WHAT WAS ONLY PARTIALLY CONFIRMED

**Disintermediation occurred, but reintermediation became more important.** The original analysis anticipated that the Internet would reduce the role of traditional intermediaries and create new cybermediaries. This was correct, but the outcome is more complex. Many traditional intermediaries lost power, while search engines, social networks, app stores, online marketplaces, payment platforms, cloud providers, logistics platforms, and advertising networks became dominant mediating infrastructures [15], [16], [17], [18].

**The idea of a simple Web presence was superseded.** The 1996 article discussed virtual stores, Web presence, content sites, virtual shopping centers, incentive sites, and search agents. These categories remain historically important, but they no longer capture contemporary digital business. Firms now operate through websites, mobile applications, social platforms, marketplaces, APIs, cloud ecosystems, generative AI interfaces, and data partnerships.

**The expected democratization of global markets was only partly achieved.** The 1996 paper argued that the Internet could allow small firms to reach international markets at lower cost [1], [19]. This remains true through e-commerce platforms, digital advertising, SaaS tools, and cross-border payment systems. However, global visibility is now mediated by ranking algorithms, advertising budgets, platform rules, reputation systems, data access, and logistics capabilities. The Internet lowered some entry barriers but created new ones linked to platform dependence and digital capabilities.

**Cost reduction was real, but cost structures were redistributed.** The 1996 article emphasized reduced search, communication, and transaction costs. This prediction was accurate at the level of information exchange, but contemporary digital operations also generate new fixed and variable costs. Firms must invest in cloud subscriptions, cybersecurity tools, data governance, integration platforms, compliance audits, AI infrastructure, specialized personnel, and digital marketing. The strategic effect is therefore not simply lower cost; it is a shift from physical transaction costs to digital coordination, control, and assurance costs.

## VI. WHAT WAS NOT FULLY ANTICIPATED

**Data became a strategic asset.** The 1996 article emphasized information flows, but it did not fully anticipate the strategic centrality of data extraction, analytics, profiling, data monetization, and algorithmic decision-making. In 2026, data is not only transmitted through the Internet; it is collected, modeled, monetized, regulated, and used to train artificial intelligence systems.

**Artificial intelligence changed the strategic meaning of the Internet.** AI-powered search, recommendation systems, chatbots, copilots, predictive analytics, and agentic automation have transformed digital competition. The Internet is now not only a channel for communication and commerce; it is also an infrastructure for intelligent automation and decision support. This extends the 1996 framing and supports the view that business strategy and digital technology strategy have become increasingly inseparable [4], [19], [20], [21], [22].

**Regulation became a defining strategic variable.** The 1996 paper identified copyright, criminality, and censorship as important constraints [1], but the regulatory landscape of 2026 is broader. Firms now operate under privacy laws, cybersecurity requirements, consumer protection rules, platform regulation, AI governance frameworks, digital services obligations, and data transfer restrictions. Regulation is no longer merely an external legal constraint; it shapes product design, data architecture, market access, and competitive positioning.

**Digital sustainability and sovereignty became strategic concerns.** The 1996 article briefly addressed environmental and physical dimensions, but it did not foresee the scale of energy consumption, e-waste, data center expansion, dependence on global cloud infrastructure, submarine cables, semiconductors, and geopolitical concerns over digital sovereignty. By 2026, data centers, AI workloads, cloud infrastructure, and edge computing raise energy, resilience, dependency, and governance challenges that directly affect competitiveness and strategic autonomy [12], [23], [24].
**Cloud concentration became a new dependency layer.** The 1996 Internet was described as a decentralized infrastructure. In practice, enterprise digital transformation increasingly depends on a small number of hyperscale cloud providers, software-as-a-service platforms, app stores, identity systems, and payment networks. This creates economies of scale and accelerates innovation, but it also introduces concentration risk, switching costs, jurisdictional exposure, outage dependency, and strategic lock-in. Therefore, contemporary Internet strategy must evaluate portability, interoperability, exit options, service-level dependency, and vendor concentration as part of competitive analysis [11], [12].
**AI agents extend Internet strategy from communication to autonomous coordination.** AI-enabled agents, copilots, recommendation models, and automated workflows increasingly act on behalf of users and organizations. This adds a new strategic layer because competitive advantage depends on how firms govern machine-generated actions, model outputs, training data, permissions, audit trails, and accountability. The Internet becomes not only a space for human-to-human and firm-to-customer interaction, but also a substrate for machine-to-machine coordination. This makes governance of non-human actors, model risk, and algorithmic transparency central to enterprise strategy [13], [14].
**Digital sovereignty reframes competitiveness as control over the digital stack.** In 1996, strategic control was mainly discussed through market positioning and information access. By 2026, control also concerns cloud jurisdiction, data residency, model governance, operational continuity, cybersecurity assurance, and the ability to move workloads across providers. Sovereignty is therefore not only a public policy concern; it is an enterprise capability related to resilience, compliance, and strategic autonomy. Firms that cannot explain where data is processed, who controls critical infrastructure, and how AI systems are governed face higher regulatory and operational risk [12], [25].

## VII. CRITICAL COMPARISON: 1996 EXPECTATIONS AND 2026 REALITY

The comparison between the 1996 argument and the 2026 digital environment should not be read as a simple checklist of correct or incorrect predictions.

**TABLE I**
CRITICAL COMPARISON OF 1996 INTERNET STRATEGY PROPOSITIONS AND 2026 REALITIES

| 1996 proposition | 2026 status | Strategic meaning |
|---|---|---|
| **Internet supports competitive advantage.** | Confirmed | From connectivity to digital operating infrastructure. |
| **Electronic markets will grow.** | Confirmed | From online stores to marketplaces, platforms, and digital payments. |
| **Cybermediaries will emerge.** | Strongly confirmed | New intermediaries control visibility, trust, payment, and access. |
| **Transaction and communication costs will fall.** | Partly confirmed | Lower coordination costs, but higher costs for security, compliance, cloud, and data. |
| **Security and payments are strategic constraints.** | Confirmed and amplified | From payment trust to enterprise-wide cyber resilience. |
| **Disintermediation will reshape sectors.** | Transformed | Disintermediation was followed by platform reintermediation. |
| **Web presence is strategically important.** | Confirmed but outdated | Digital presence now spans websites, apps, APIs, platforms, and AI interfaces. |
| **Internet use requires internal rules.** | Confirmed | Rules evolved into cybersecurity, data, AI, vendor, and compliance governance. |

The more important issue is how the same strategic concepts changed scale, scope, and technical meaning over three decades. In 1996, the Internet was framed mainly as a communication, information, and transaction infrastructure. By 2026, it functions as a layered enterprise environment composed of cloud platforms, data ecosystems, algorithmic services, cybersecurity controls, payment infrastructures, digital identity systems, and regulatory obligations.

This change affects the interpretation of the original paper's propositions in three ways. First, several predictions were confirmed because connectivity, electronic markets, cybermediaries, and online communities became routine components of enterprise strategy. Second, some predictions were transformed because disintermediation did not eliminate mediation; instead, it produced platform-based reintermediation in which a smaller number of firms control visibility, access, trust, payment, and distribution. Third, important new layers emerged that were not central in the 1996 article, particularly data governance, AI-enabled automation, cloud dependency, cybersecurity resilience, and digital sovereignty.

Table I synthesizes these three patterns by linking each original proposition to its 2026 status and underlying strategic mechanism. The comparison indicates that the 1996 article was strongest in anticipating structural and value-chain effects, but less complete in accounting for the institutional, infrastructural, and algorithmic dimensions that now shape Internet-based competition.

Table I should therefore be read as a compact synthesis of the detailed analysis developed in the preceding sections. The first group of propositions was clearly confirmed: the Internet became a foundation for competitive advantage, electronic markets expanded into routine commercial infrastructure, cybermediaries became central actors, and security became a strategic condition for digital business. These confirmations show that the 1996 article accurately identified the direction of change: connectivity would affect industry structure, market access, value-chain activities, and organizational coordination. However, the mechanisms through which these effects occurred became more complex than the original framing suggested.

The most important transformation concerns intermediation. Although some traditional intermediaries lost relevance, the dominant pattern was not simple disintermediation. Instead, platform-based reintermediation reorganized competition around search engines, marketplaces, app stores, payment providers, social networks, cloud platforms, and logistics infrastructures. These actors became strategic gatekeepers because they control ranking, reputation, data access, pricing rules, identity management, payment trust, and customer visibility. Therefore, the Internet did not remove mediation; it shifted mediation to technically sophisticated and highly concentrated digital infrastructures.

The table also highlights the need to reinterpret cost reduction. Communication, search, and transaction costs decreased substantially, but firms simultaneously acquired new digital costs related to cybersecurity, compliance, cloud services, data governance, systems integration, AI infrastructure, and digital talent. In this sense, the strategic effect of the Internet is better understood as cost redistribution and capability reconfiguration rather than simple cost reduction. Similarly, the idea of a Web presence must now be replaced by a broader concept of digital presence, including websites, mobile applications, APIs, online communities, marketplaces, AI interfaces, and ecosystem participation.

## VIII. IMPLICATIONS FOR ENTERPRISE STRATEGY IN 2026

The principal implication is that the Internet can no longer be managed as a separate technological channel. Digital business strategy increasingly integrates business and technology choices [4], while digital innovation depends on distributed combinations of platforms, data, and organizational capabilities [6]. Accordingly, the progression from connectivity to AI-enabled enterprise systems requires coordinated decisions about digital infrastructure, data capabilities, AI-supported processes, cybersecurity, regulatory compliance, platform participation, customer experience, and organizational learning. Porter's value-chain logic remains relevant for identifying where technology affects cost and differentiation [7], but it must now be extended to account for ecosystem interdependence, data flows, algorithmic mediation, and platform governance [15], [17].

For Portugal, the contrast between 1996 and 2026 is especially significant. The original article examined a small, emerging Internet market in which domain registration, access provision, and hosting capacity were still limited. By 2026, e-commerce had become an established component of the national business environment [10], and the policy emphasis had shifted toward digital maturity, cloud adoption, AI capability, cybersecurity readiness, and firms' effective use of advanced digital services [26]. The strategic problem is therefore no longer access alone, but whether firms can convert available digital infrastructure into productivity, innovation, resilience, and international competitiveness.

An enterprise strategy for 2026 should incorporate four design principles. First, architectural modularity should allow applications, data stores, APIs, and AI services to evolve without excessive technological lock-in; platform architecture and governance must therefore be treated as joint strategic choices [17]. Second, governance-by-design should embed privacy, cybersecurity, access control, compliance, and model-risk safeguards in systems before deployment, especially as AI agents expand the number of non-human actors operating across organizational workflows [13], [14]. Third, firms require an explicit ecosystem strategy because value creation depends on complementors, platform rules, cloud providers, data partners, and developer communities [15], [19]. Fourth, resilience planning should combine redundancy, incident response, workload portability, vendor exit options, and continuity arrangements for critical digital services [11], [12], [24].

These principles change how the value chain should be applied. Procurement now includes governance of cloud, software, data, and AI suppliers, not only price negotiation [11], [17]. Operations increasingly depend on data pipelines, automation, and AI-supported workflows [20], [21]. Marketing and sales are shaped by algorithmic visibility, recommendation systems, customer-data platforms, and platform access [15], while service activities increasingly use remote diagnostics, conversational agents, and community knowledge bases [8]. Firm infrastructure must encompass enterprise architecture, cybersecurity, data stewardship, and regulatory intelligence; human resource management must develop digital skills, AI literacy, responsible-use practices, and organizational capacity for adaptation [19], [24]. The strategic unit of analysis is therefore no longer an isolated internal activity, but the configuration of interdependent capabilities across the firm and its digital ecosystem.

### *A. Research Challenge and Technological Outlook*

**Main research challenges.** Research must explain when AI-enabled digital infrastructures generate durable strategic advantage, when they instead deepen dependence on platforms, cloud providers, foundation models, and proprietary data ecosystems, and which governance arrangements determine these outcomes. Platform dependence can restrict control over architecture, complementors, and value appropriation [17], while cloud concentration and jurisdictional exposure create additional risks for resilience and strategic autonomy [11], [12], [25]. AI agents intensify these tensions because firms must coordinate human and machine agency while preserving decision rights, data provenance, accountability, and effective oversight [13], [14]. The central theoretical problem is therefore the relationship between capability amplification and control erosion: architectures that increase speed, scale, and automation may also produce opacity, lock-in, systemic exposure, and weakened organizational autonomy. Research should determine how modularity, data stewardship, human oversight, model assurance, interoperability, and credible exit options interact across firm, ecosystem, and institutional levels to strengthen - or undermine - resilience and sustained competitive advantage [30].

**Future research agenda.** Four connected streams merit priority. First, research on *AI-enabled coordination and firm boundaries* should test whether agent-mediated workflows internalize activities, expand outsourcing, or generate hybrid arrangements between firms and platforms; this stream extends current work on AI as a core technology of digital transformation [21]. Second, research on *strategic AI resources* should distinguish access to general-purpose models from value created through proprietary data quality, process integration, complementary skills, and organizational learning [19], [20]. Third, research on *governance and resilience* should examine whether modular architectures, sovereign-cloud arrangements, open standards, domain-specific models, and federated or edge systems reduce concentration and cyber risk without unacceptable losses in performance or scale [11] - [14], [25], [30]. Fourth, research on *institutional and sustainability constraints* should assess how regulation, energy availability, data residency, and digital-sovereignty requirements shape architecture, location decisions, market entry, and innovation [23], [24]. Across these streams, studies should explain variation in productivity, innovation, strategic autonomy, resilience, accountability, and value appropriation rather than treating adoption as the principal outcome [31], [32].

**Empirical priorities.** This agenda requires research designs capable of identifying mechanisms and causal relationships, including longitudinal firm-level panels, cross-country comparisons, ecosystem network analysis, matched case studies, and field experiments with bounded AI deployments. Three propositions offer an initial basis for cumulative testing. First, AI-enabled coordination should improve performance only when decision rights, data provenance, monitoring, and escalation mechanisms are aligned, consistent with the importance of process integration in digital transformation [21]. Second, architectural modularity and credible switching options should weaken the adverse effect of platform and cloud dependence on strategic autonomy [11], [12], [17]. Third, the performance value of AI should be mediated by process redesign, complementary capabilities, and organizational learning rather than by model access alone [19], [20]. Tests of these propositions should examine boundary conditions such as firm size, sectoral regulation, task uncertainty, data sensitivity, and national infrastructure capacity, while explicitly evaluating trade-offs among efficiency, resilience, sovereignty, and accountability [24], [25], [30]. Such designs would move research beyond descriptive accounts of adoption toward explanatory evidence on how hybrid human - AI systems reshape enterprise strategy [31], [32].

## IX. CONCLUSION

The 1996 article was highly forward-looking. Its central claim - that the Internet would affect enterprise strategy through environmental analysis, industry structure, value-chain activities, marketing, distribution, communities, and security - has been substantially confirmed. The original analysis correctly identified many mechanisms through which the Internet would transform firms, even though the scale and institutional consequences of that transformation later became far broader.

However, the reality of 2026 is more complex than the 1996 framework could fully anticipate. Over three decades, the Internet evolved from a communication and commerce infrastructure into a platformized, data-driven, cloud-dependent, regulated, and geopolitically significant environment in which AI increasingly mediates decisions, interactions, and operations. AI is therefore not a replacement for the paper's Internet-strategy perspective but the latest extension of it. The central conceptual update is that Internet strategy can no longer be treated as a subset of business strategy; for many firms, business strategy is inseparable from the digital infrastructures and AI-enabled capabilities through which it is executed [4], [27]. Future research should examine how organizations build sustainable advantages in digital ecosystems characterized by platform dependency, data asymmetry, AI automation, cybersecurity risk, regulatory pressure, and rapid technological change.